\documentclass[12pt]{article}
\usepackage{amsmath, amssymb}
\usepackage{authblk}        
\usepackage{changepage}     
\usepackage[margin=1in]{geometry}       
\usepackage{setspace}       
\usepackage{hyperref}
\usepackage{graphicx}
\usepackage{amsfonts}
\usepackage{hyperref}
\usepackage{braket}
\usepackage{stackrel}
\usepackage{diagbox}
\usepackage[multiple]{footmisc}

\title{Dark states in an integrable $XY$ central spin model}

\author[1]{Jaco van Tonder\thanks{Email: \href{mailto:w.vantonder@student.uq.edu.au}{w.vantonder@student.uq.edu.au}}}
\author[1]{Jon Links}
\affil[1]{School of Mathematics and Physics, The University of Queensland, 4072, Australia}

\date{}  

\begin{document}
\maketitle

\begin{abstract}
Eigenstates of central spin models in which the central spin is unentangled with the environment are known as dark states. They have recently been observed in a class of integrable $XX$ models. Here we find that dark states are present in $XY$ models, but only for particular configurations of the central spin magnetic field. We show this via an explicit construction of the states. 
\end{abstract}

%
%
%
%
%

\section{Introduction}\label{sec:Introduction}

Central spin models have gained renewed attention due to their possible applications in modern quantum technologies focused on quantum sensing and metrology \cite{slcwpl14}. This is due to their integrability allowing high-fidelity control of these systems at a mesoscopic scale, where the exponentially increasing size of the Hilbert space would in general be prohibitive \cite{dldwlrd19}. The central spin allows the dynamics of the spin bath to be monitored, and for feedback to be used to steer the dynamics of the bath in a desired direction. There are several physical or engineered systems for which central spin models provide a theoretical model. These include nitrogen vacancy (NV) centres in diamonds \cite{yjgmgcl12}, room temperature quantum memory storage \cite{til03, vcc20, vcppc20}, quantum batteries \cite{lsswy21} and quantum dots \cite{cbb07, umakvm13} -- an active area of research which gained public attention through the 2023 Nobel prize for Chemistry \cite{Sanderson2023}.

On the theoretical side, these models have been known for their integrability since Gaudin's seminal paper \cite{g76}. Integrability allows for the analytic solution of the eigenstates and eigenspectrum of the models through the Bethe Ansatz. This makes these models well suited for studies of their equilibrium and dynamical behaviour; in particular, tests of the Eigenstate Thermalisation Hypothesis (ETH) and investigation of many-body localisation \cite{a15, l11, tlpcc23, vcppc20}. Integrability is also essential for perturbative solutions of physical models. For example, integrability provides a wavefunction Ansatz in variational eigensolvers to model strong electron correlation in quantum chemistry \cite{fbkj20}.

Integrability is also of great interest in mathematical physics. There is ongoing work on extending or modifying known integrable models to obtain new ones. This has recently led to establishing that the $XX$ model, which models certain resonant dipolar spin systems, is integrable \cite{df22, vcc20}. Integrability was subsequently extended to the $XY$ model \cite{vl23} using conserved charges discussed in \cite{sil20}, and the same charges with self-interaction obtained by Skrypnyk \cite{s22}. These charges were diagonalised using several modifications of the standard Algebraic Bethe Ansatz for arbitrary spin \cite{s22}. These expressions for the eigenvalues can also be obtained through the use of functional relations as described in \cite{vl23}. 

Here we derive the $XY$ central spin model Hamiltonian and its charges through a limit of the charges for the $XYZ$ Richardson-Gaudin model in \cite{s22}. We employ a reparametrisation of the results of  \cite{s22} to obtain regularised eigenvalue and eigenstate results for the central spin model which reproduce the diagonalisation results in the literature \cite{cdv16,vcc20} in the appropriate limits. It is found that dark states, states in which the central spin is not entangled with the environment, are present for special configurations of the magnetic field $\vec{B} = (B^x, B^y, B^z)$. This is reminiscent of the $XX$ model for which these occur for an out-of-plane magnetic field, $B^x = B^y = 0$ \cite{df22, vcc20}.

In Sect. \ref{sec: Overview} we give a summary of the results found in this work. In the following Sect. \ref{sec: Derivation of two classes} we show how the $XY$ central spin Hamiltonian and its conserved charges for spin-$1/2$ central spin can be derived from the XYZ Gaudin model charges. This allows the eigenstate results of \cite{s22} to be used to find the eigenstates for the $XY$ central spin model in Sect. \ref{sec: Eigenstates}. In Sec. \ref{sec: Dark states} we show that for special magnetic field configurations dark states occur, in analogy to those that were seen in the out-of-plane $XX$ model \cite{vcc20}. Concluding remarks are given in Sect. \ref{sec: Conclusion}. In the Appendix \ref{sec: Appendix} we take the isotropic limit to confirm that these recover the results for the $XX$ model. This is supplemented with some heuristic arguments and numerical results about how the dark states emerge from the states with generic magnetic field configurations.

\section{Overview of main results}\label{sec: Overview}

Consider a set of $L+1$ spin operators $\{S^x_j,\,S_j^y,\,S^z_j: j=0,1,\dots, L\}$ satisfying the standard canonical commutation relations
\begin{align}\label{Spin operators}
	&[ S^\theta_j,\,S^\eta_k]=i\delta_{jk} \sum_{\kappa\in\{x,y,z\}}\varepsilon^{\theta\eta\kappa} S^\kappa_j ,
\end{align}
where $i = \sqrt{-1}$ and $\varepsilon^{\theta\eta\kappa}$ is the Levi-Civita symbol.
The main focus of this work is the central spin Hamiltonian \cite{vl23}
\begin{align}\label{H}
	H &= B^xS^x_0+B^yS^y_0+B^z S^z_0+2\sum_{j = 1}^{L}(	X_j S^x_0S^x_j+Y_j S^y_0S^y_j)
\end{align}
describing a central spin-$1/2$ particle interacting with a bath of $L$ surrounding spins through an inhomogeneous $XY$ interaction. The central spin is also subject to an external magnetic field $\vec{B} = \{B^x, B^y, B^z\}$. In \cite{vl23} it was shown that the model is integrable when $Y_j^2 - X_j^2 = 2\beta$, with the constant $\beta \in \mathbb R$ being free to be chosen through rescalings of the parameters and magnetic field components. The spins $s_j$ of the surrounding bath particles are arbitrary.

Along with $\beta$ we introduce a set of distinct parameters $\{\epsilon_j:j=1,\dots,L\}$ such that $\epsilon_j\pm\beta>0$ for all $j=1,\dots,L$. We also define $f_j^\pm =\sqrt{\epsilon_j\pm\beta}$. In terms of these it is convenient to use the parametrisation
\begin{align*}
	X_j = \sqrt{\epsilon_j - \beta}, \quad Y_j = \sqrt{\epsilon_j + \beta}.
\end{align*} 
 The Hamiltonian (\ref{H}) possesses $L$ conserved charges \cite{vl23}
\begin{align*}\label{Q_I}
	Q_i &= -2 S_0^z S_i^z+\frac{B^x}{f_i^+}S_i^x+\frac{B^y}{f_i^-}S^y_i
	+\frac{f_i^-}{f_i^+}S_i^xS_i^x+\frac{f_i^+}{f_i^-}S_i^y S_i^y \nonumber\\
	&\quad+2\sum_{\stackrel[j\neq i]{}{j=1}}^{L}\frac{1}{\epsilon_i-\epsilon_j}(f_i^+f_j^-S^x_iS^x_j+f_i^-f_j^+S^y_iS^y_j
	+f_j^+f_j^-S^z_iS^z_j).
\end{align*}
To describe its eigenstates we note that we can also bring the Hamiltonian into the form
\begin{align*}
	H &= B^z S_0^z+S_0^{-}A^{+}+S_0^{+}A^{-},
\end{align*}
where
\begin{equation}\label{Class I supercharges}
	A^{\pm} = (B^x\pm iB^y)I/2+\sum_{j=1}^{L}(f_j^-S^x_j\pm if_j^+ S^y_j).
\end{equation}

Along with the eigenstates $\ket{\pm}_0$ of $S^z_0$, define reference states from which to build the eigenstates via the Bethe Ansatz
\begin{align*}\label{Reference state}
	v^{\pm}_B & = v_{1}^{\pm}\otimes\cdots\otimes v_{L}^{\pm}.
\end{align*}
Here the $v_k^{\pm}$ satisfy
\begin{align*}
	T_{k}^{z}v_{k}^{+}=-s_{k}v_{k}^{+},\; T_{k}^{-}v_{k}^{+}=0; \quad
	T_{k}^{z}v_{k}^{-}=s_{k}v_{k}^{-},\; T_{k}^{+}v_{k}^{-}=0
\end{align*}
for the operators
\begin{align*}
	T_{k}^{z} & =-i\frac{\sqrt{2\beta}}{f_{k}^{-}}S_{k}^{x}-\frac{f_{k}^{+}}{f_{k}^{-}}S_{k}^{z}, \quad	T_{k}^{\pm} & =\pm iS_{k}^{y}-\frac{f_{k}^{+}}{f_{k}^{-}}S_{k}^{x}+i\frac{\sqrt{2\beta}}{f_{k}^{-}}S_{k}^{z}.
\end{align*}
In terms of the operators
\begin{align*}
	S^{\pm}_k(u) & = \pm(2k-1)i\sqrt{2\beta}+B^x \pm i B^y-i \zeta \sqrt{2\beta}\\
	&\quad +2\sum_{j=1}^{L}\frac{f^+_j (u-\beta)}{\epsilon_{j}-u}T_{j}^{\mp}-2i\sqrt{2\beta}\sum_{j=1}^{L}T_{j}^{z},
\end{align*}
the eigenstates generically take the form
\begin{align*}
	\ket{\kappa; u_1,\dots,u_{M}} & =\left.\left[A^{+} - \kappa S_{0}^{+}\right]\ket{-}_0 \otimes S^{+}_1(u_{1})S^{+}_2(u_{2})\cdots S^{+}_{M}(u_{M})v^{-}_B\right|_{\zeta =-1}
\end{align*}
for $M = N_s \equiv \sum_{j=1}^L 2s_j$ and have energy $E  = -B^z/2-\kappa$. They depend on a parameter $\kappa$ satisfying
\begin{align*}
	& \kappa^{2}+B^z\kappa+i B^x\sqrt{2\beta}-2\beta-2\sum_{j=1}^{L}(\epsilon_{j}-\beta)s_{j}+2\sum_{m=1}^{M}(u_{m}-\beta)\\
	& =-\frac{1}{4}\left(\left((M+2)\sqrt{2\beta}- iB^x\right)^2-(B^y)^2\right)\frac{\prod_{m=1}^{M}(u_{m}-\beta)}{\prod_{j=1}^{L}(\epsilon_{j}-\beta)^{2s_{j}}}.
\end{align*}
and a set of Bethe roots $\{u_m\}_{m=1}^M$ satisfying the Bethe Ansatz equations
\begin{align*}
	&2u_{l}+4\beta - i B^x\sqrt{2\beta}+2\sum_{j=1}^{L}\frac{(\epsilon_{j}-\beta)(u_{l}+\beta)}{u_{l}-\epsilon_{j}}s_{j}-2\sum_{m\neq l}^{M}\frac{(u_{m}-\beta)(u_{l}+\beta)}{u_{l}-u_{m}}\nonumber\\
	&\;= \frac{1}{4}\left(\left((M+2)\sqrt{2\beta} - i B^x\right)^2-(B^y)^2\right) \prod_{j=1}^{L}\left(\frac{\epsilon_{j}-u_{l}}{\epsilon_{j}-\beta}\right)^{2s_{j}}\prod_{m\neq l}^{M}\frac{u_{m}-\beta}{u_{m}-u_{l}}.
\end{align*}

These states are generically bright states, in other words entangled states of the bath spins and the central spin. However, if the magnetic field components satisfy
\begin{equation}\label{eq: Dark state magnetic field config}
	B^x = 0, \quad \frac{B^y}{2\sqrt{2\beta}}  = M' - N_s/2
\end{equation}
for $M' \in \{0, 1, \dots, N_s\}$, then some of these states become dark states in which the eigenstates are product states of the central spin and bath spins. These are the eigenstates
\begin{align*}
	&\left.\ket{+}_0 \otimes S^{-}_1(u_{1})\cdots S^{-}_{M'}(u_{M'})\, v^{+}_B\right|_{\zeta = - 1}, \\
	&\left.\ket{-}_0 \otimes S^{+}_1(u_{1})\cdots S^{+}_{M'}(u_{M'})\, v^{-}_B\right|_{\zeta = + 1}.
\end{align*}
These have respective energy $E = \pm B^z/2$ and Bethe roots satisfying the Bethe Ansatz equations
\begin{align*}
	2\beta+2\sum_{j=1}^{L}\frac{(u_{l}+\beta)(\epsilon_{j}-\beta)}{u_{l}-\epsilon_{j}}s_{j}-2\sum_{m\neq l}^{M'}\frac{(u_{l}+\beta)(u_{m}-\beta)}{u_{l}-u_{m}}=0.
\end{align*}

\section{Derivation of integrable $XY$ central spin Hamiltonians}\label{sec: Derivation of two classes}
A class of integrable central spin models, with spin-1/2 central spin and arbitrary bath spins, is derived as follows.  Introducing an additional parameter $\epsilon_0$ we consider the $L+1$ conserved charges\footnote{We use $\zeta$ instead of $B^z$ to prevent confusion with $B^z$ of the $XY$ Hamiltonian (\ref{H_I}) below.} \cite{s22,vl23}
\begin{align}\label{QQ_i}
	&\mathcal{Q}_i = \zeta S^z_i+\frac{B^x}{f_i^+}S^x_i+\frac{B^y}{f_i^-}S^y_i+\frac{f_i^-}{f_i^+}S^x_iS^x_i+\frac{f_i^+}{f_i^-}S^y_iS^y_i \nonumber\\
	&\quad\quad\quad\quad+2\sum_{j\neq i}^{L}\frac{1}{\epsilon_i-\epsilon_j}(f_i^+f_j^-S^x_iS^x_j+f_i^-f_j^+S^y_iS^y_j
	+f_j^+f_j^-S^z_iS^z_j), \\
	&[\mathcal{Q}_i, \mathcal{Q}_j] = 0,\quad\forall {i,j}\in\{0,1,\dots,L\}\nonumber.
\end{align}
We take $\{S^x_0,\,S_0^y,\,S^z_0\}$ to be the spin-$1/2$ operators, identify the zero subscript with the central spin and the other subscripts with the bath spins. The zeroth charge $\mathcal{Q}_0$ will then become the central spin Hamiltonian and $\{\mathcal {Q}_j:j=1,\dots,L\}$ its conserved charges. Noting that for spin-1/2 operators we have 
\begin{align*}
	\left(S^x_0\right)^2=\left(S^y_0\right)^2=\frac{1}{4} I,
\end{align*}
it is convenient for the zeroth charge to subtract off the squared spin operator terms. After making the change of variables $\{\zeta \mapsto B^z/\sqrt{\epsilon_0}\}$ for the magnetic field, we obtain the XY central spin model through the limit as $\epsilon_0 \to \infty$ with the Hamiltonian being
\begin{align}\label{H_I}
	H &= \lim\limits_{\epsilon_0\to \infty} \left[ \sqrt{\epsilon_0}\mathcal{Q}_0 -\frac{\sqrt{\epsilon_0}}{4}\left(\frac{f^-_0}{f^+_0}+\frac{f^+_0}{f^-_0}\right)\right]\nonumber\\
	&= B^xS^x_0+B^yS^y_0+B^z S^z_0+2\sum_{j = 1}^{L}(f_j^-S^x_0S^x_j+f_j^+S^y_0S^y_j),
\end{align}
and the charges
\begin{align}\label{Q_I}
	Q_i &= \lim\limits_{\epsilon_0\to \infty} \mathcal{Q}_i \nonumber\\
	&= -2 S_0^z S_i^z+\frac{B^x}{f_i^+}S_i^x+\frac{B^y}{f_i^-}S^y_i
	+\frac{f_i^-}{f_i^+}S_i^xS_i^x+\frac{f_i^+}{f_i^-}S_i^y S_i^y \nonumber\\
	&\quad+2\sum_{\stackrel[j\neq i]{}{j=1}}^{L}\frac{1}{\epsilon_i-\epsilon_j}(f_i^+f_j^-S^x_iS^x_j+f_i^-f_j^+S^y_iS^y_j
	+f_j^+f_j^-S^z_iS^z_j),
\end{align}
reproducing those introduced in \cite{vl23}.

\section{The Bethe Ansatz} \label{sec: Eigenstates}

The eigenstates of the charges $\mathcal{Q}_j$, as well as their eigenvalues, have been found in  \cite{s22}. Here we recall the algebraic setup and the results required in order to obtain the eigenstates and eigenvalues of $H$ and its conserved charges $Q_j$.

\subsection{Relations for the Algebraic Bethe Ansatz}\label{sec: Eigenstate setup}

Introduce the Lax algebra generators
\begin{align*}
	S^{-}(u) & = B^x-i B^y-i \zeta \sqrt{2\beta}+2\sum_{j=0}^{L}\frac{f^+_j (u-\beta)}{\epsilon_{j}-u}T_{j}^{+}-2i\sqrt{2\beta}\sum_{j=0}^{L}T_{j}^{z}, \\
	S^{+}(u) & = B^x+i B^y-i \zeta \sqrt{2\beta}+2\sum_{j=0}^{L}\frac{f^+_j (u-\beta)}{\epsilon_{j}-u}T_{j}^{-}-2i\sqrt{2\beta}\sum_{j=0}^{L}T_{j}^{z}\\
	S^z(u) &= \sqrt{2\beta}\frac{iB^x}{u-\beta}+\zeta\frac{u+\beta}{u-\beta} +2\sum_{j=0}^{L}\frac{(u+\beta)(\epsilon_j-\beta)}{(u-\beta)(\epsilon_j-u)}T_{j}^{z}
\end{align*}
where the operators $\{T_k^{+}, T_k^{z},T_k^{-}\}$ satisfy the $\mathfrak{su}(2)$ commutation relations
\begin{align*}
	[T_k^+,T_l^-] = 2\delta_{k,l}T_k^z,\quad [T_k^z,T_l^\pm] = \pm\delta_{k,l}T_k^+,
\end{align*}
and in terms of the spin operators take the form 
\begin{align*}
	T_{k}^{z} & =-i\frac{\sqrt{2\beta}}{f_{k}^{-}}S_{k}^{x}-\frac{f_{k}^{+}}{f_{k}^{-}}S_{k}^{z}, \quad
	\\
	T_{k}^{-} & =-iS_{k}^{y}-\frac{f_{k}^{+}}{f_{k}^{-}}S_{k}^{x}+i\frac{\sqrt{2\beta}}{f_{k}^{-}}S_{k}^{z}, \quad
	T_{k}^{+} =iS_{k}^{y}-\frac{f_{k}^{+}}{f_{k}^{-}}S_{k}^{x}+i\frac{\sqrt{2\beta}}{f_{k}^{-}}S_{k}^{z}.
\end{align*}
The Lax algebra generators\footnote{Specifically these are a reparametrisation of those in \cite{s22} in order to regularise the eigenstates in the isotropic $\beta \to 0$ limit recovering the $XX$ model. See Appendix \ref{sec: Reparametrised operators} for the details.} satisfy the relations
\begin{align}\label{Lax algebra relations}
	{[}S^{\pm}(u),S^{\pm}(v){]}	&= \pm 2i\sqrt{2\beta}\left(S^\pm(v)-S^\pm(u)\right), \\
	{[}S^z(u),S^\pm(v){]}	&= \pm 2\frac{(u+\beta)(v-\beta)}{(u-v)(u-\beta)}\left(S^\pm(v)-S^\pm(u)\right), \nonumber\\
	{[}S^z(u),S^z(v){]}	&= 0, \nonumber\\
	{[}S^+(u),S^-(v){]} &= \frac{(u-\beta)(v-\beta)}{u-v}\left(S^z(v)-S^z(u)\right)\\
	&\quad\quad\quad-2i\sqrt{2\beta}S^-(v)-2i\sqrt{2\beta}S^+(u). \nonumber
\end{align}
We also define operators shifted by a constant term, necessary to build the appropriate Bethe states,
\begin{align*}
	S^{\pm}_k(u) = S^{\pm}(u)\pm(2k-1)i\sqrt{2\beta}I, \quad k\in \mathbb{Z}_{\geq 0}
\end{align*}
and with these define
\begin{align}\label{eq: Bethe state operator}
	\mathcal{S}^{\pm}(u_1, u_2, \dots, u_M) = S^{\pm}_1(u_1) S^{\pm}_2(u_2) \cdots S^{\pm}_M(u_M).
\end{align}
Following the notation of \cite{s22} we also introduce the reference states $v^{\pm}$ defined as
\begin{align}\label{Reference state}
	v^{\pm} & = v_{0}^{\pm} \otimes v_{1}^{\pm}\otimes\cdots\otimes v_{L}^{\pm}
\end{align}
where the $v_k^{\pm}$ are respectively lowest and highest weight states satisfying
\begin{align*}
	T_{k}^{z}v_{k}^{+}=-s_{k}v_{k}^{+},\; & T_{k}^{-}v_{k}^{+}=0,\\
	T_{k}^{z}v_{k}^{-}=s_{k}v_{k}^{-},\; & T_{k}^{+}v_{k}^{-}=0
\end{align*}
for half non-negative integers $s_k$, i.e. $s_k \in \{0, 1/2, 1, 3/2, \dots\}$. For the spin labelled by the 0 subscript, for which $s_0 = 1/2$, we use the following notational convention
$$
	\ket{+}_0 = v^-_0, \quad \ket{-}_0 = v^+_0.
$$
For conciseness we will also use the notation
$$
	v_B^\pm =  v_{1}^{\pm}\otimes\cdots\otimes v_{L}^{\pm}.
$$

\subsection{Generic eigenstates of the charges $\mathcal{Q}_j$}\label{subsec: Eigenstates of H}

To describe the eigenstates of the $XY$ model for generic magnetic field configurations we will use the corresponding results of \cite{s22} for the charges $\mathcal{Q}_j$. 

As was shown in \cite{s22}, the charges
\begin{align*}
	&\mathcal{Q}_j = \zeta S^z_j+\frac{B^x}{f_j^+}S^x_j+\frac{B^y}{f_j^-}S^y_j+\frac{f_j^-}{f_j^+}S^x_jS^x_j+\frac{f_j^+}{f_j^-}S^y_jS^y_j \nonumber\\
	&\quad\quad\quad\quad+2\sum_{k\neq j}^{L}\frac{1}{\epsilon_j-\epsilon_k}(f_j^+f_k^-S^x_jS^x_k+f_j^-f_k^+S^y_jS^y_k
	+f_k^+f_k^-S^z_jS^z_k),
\end{align*}
have the eigenstates constructed from raising operators
\begin{align}\label{eq: Bethe states}
	\ket{u_1,\dots,u_M} & =\mathcal{S}^{+}(u_{1},\dots,u_{M})v^{-},
\end{align}
with respective eigenvalues
\begin{align}
	\mathfrak{q}_{j} & = \frac{f_{j}^{+}}{f_{j}^{-}}s_{j}(s_{j}+1)-\frac{\epsilon_{j}+3\beta}{f_{j}^+f_j^-}s_{j}^{2}-\left(iB^x\frac{\sqrt{2\beta}}{f_{j}^{+}f_j^-}+\zeta \frac{f_{j}^{+}}{f_j^-}\right)s_{j}\nonumber\\
	&\quad+2\frac{f_{j}^{+}}{f_j^-}\sum_{m=1}^{M}\frac{u_{m}-\beta}{u_{m}-\epsilon_{j}}s_{j}+2\frac{f_{j}^{+}}{f_j^-}\sum_{k\neq j}^{L}\frac{\epsilon_{k}-\beta}{\epsilon_{j}-\epsilon_{k}}s_{j}s_{k}.
	\label{eq:non-singular eigenvalues}
\end{align}
The Bethe roots $\{{u_m}\}_{m=1}^M$, with $M = 1+N_s$ where $N_s = \sum_{j=1}^L 2s_j$, are the solutions to the set of Bethe Ansatz equations
\begin{align}\label{eq: BAE for spin chain charges}
	& u_{l}+3\beta-iB^x\sqrt{2\beta}-\zeta(u_{l}+\beta) \nonumber\\
	&+2\sum_{j=0}^{L}\frac{(\epsilon_{j}-\beta)(u_{l}+\beta)}{u_{l}-\epsilon_{j}}s_{j}-2\sum_{m\neq l}^{M}\frac{(u_{m}-\beta)(u_{l}+\beta)}{u_{l}-u_{m}} \nonumber\\
	& =\frac{1}{4}\left(\left((M+1)\sqrt{2\beta}-(iB^x+\zeta\sqrt{2\beta})\right)^2-(B^y)^2\right) \nonumber\\
	&\quad\times \prod_{j = 0}^{L}\left(\frac{\epsilon_{j}-u_{l}}{\epsilon_{j}-\beta}\right)^{2s_{j}}\prod_{m\neq l}^{M}\frac{u_{m}-\beta}{u_{m}-u_{l}}.
\end{align}

It was also found in \cite{s22} that some of the eigenstates reduce to
\begin{align}\label{eq: irr Bethe states}
	\ket{u_1,\dots,u_{M'}} & =\mathcal{S}^{+}(u_{1},\dots,u_{M'})v^{-},
\end{align}
with the $M' \in \{0, 1, \dots, N_s\}$ Bethe roots satisfying ``standard" Bethe Ansatz equations
\begin{align}\label{eq: standard BAE for spin chain charges}
	& u_{l}+3\beta-iB^x\sqrt{2\beta}-\zeta(u_{l}+\beta) \nonumber\\
	&+2\sum_{j=0}^{L}\frac{(\epsilon_{j}-\beta)(u_{l}+\beta)}{u_{l}-\epsilon_{j}}s_{j}-2\sum_{m\neq l}^{M'}\frac{(u_{m}-\beta)(u_{l}+\beta)}{u_{l}-u_{m}} = 0
\end{align}
when the magnetic field satisfies
 \begin{align}\label{eq: B field cond}
	\frac{-i B^x - B^y}{\sqrt{2\beta}}-\zeta & = N_{s}-2M'.
\end{align}
In other words, one can see (\ref{eq: irr Bethe states}) as a generic Bethe state (\ref{eq: Bethe states}) in which $N_s - M'$ of the roots become equal to $\beta$. Assuming that $B^x, B^y, \zeta$ are real valued, as required for the charges to be Hermitian, imposes $B^x = 0$.

Before moving onto the eigenstates of the central spin model, we note that for generic model parameters the states can be constructed by either applying the lowering operators to the highest weight state $v^+$, or the raising operators to the lowest weight state $v^-$, with each separately constituting a complete set of eigenstates. Completeness follows from quadratic identities satisfied by the charges $\mathcal{Q}_j$ similar to those seen in \cite{vl23}, and extended to the eigenvalues. From these the Bethe Ansatz equations are seen to be the consistency equations for the parametrisation in terms of the roots $\{u_m\}$. Completeness is then argued similarly as in \cite{l17b,l17c} by assuming regularisability and the one-dimensionality of the simultaneous eigenspace for generic parameters. 

\subsection{Generic eigenstates of the $XY$ model}\label{sec: generic eigenstates}
We now proceed to obtain the eigenstates of the $XY$ model with the limits defined in Sect. $\ref{sec: Derivation of two classes}$. Recall from (\ref{H_I}) that we obtain the Hamiltonian by taking the limit of $\mathcal{Q}_0$ as $\epsilon_0 \to \infty$ after applying a rescaling and a constant shift.
Hence the eigenstates are found from the eigenstates of the charge $\mathcal{Q}_0$ in the above limit. Due to the possibility that the Bethe roots $\{u_m\}$ diverge, we analyse the behaviour as two cases.

\subsubsection*{Finite roots in the limit.}

Assuming the roots stay finite in the limit, the raising operators
become
\begin{align*}
	\tilde{S}^{+}(u) & =B^x + iB^y +2i\sqrt{2\beta}S_{0}^{z}\\
	&\quad+2\sum_{j=1}^{L}\frac{f_{j}^{+}(u-\beta)}{\epsilon_{j}-u}\left( -iS_{j}^{y}-\frac{f_{j}^{+}}{f_{j}^{-}}S_{j}^{x}+i\frac{\sqrt{2\beta}}{f_{j}^{-}}S_{j}^{z}\right)\\
	& \quad+2\sqrt{2\beta}\sum_{j=1}^{L}\left(-\frac{\sqrt{2\beta}}{f_{j}^{-}}S_{j}^{x}+i\frac{f_{j}^{+}}{f_{j}^{-}}S_{j}^{z}\right)
\end{align*}
Hence the eigenstates take the form of product states of the central spin and bath spins
\begin{align*}
	\ket{u_1,\dots,u_M} 
	& = \left.\ket{-}_0\otimes \mathcal{S}^{+}(u_{1},u_{2},\dots,u_{M})v_B^{-}\right|_{\zeta = +1},
\end{align*}
namely dark states. However, we will show in Sec. \ref{sec: Dark state conditions} that dark states only occur when (\ref{eq: Dark state magnetic field config}) holds. In this case the Bethe Ansatz has to be modified to that of Sec. \ref{sec: Dark states}. 

\subsubsection*{One of the roots diverges.}

Assuming that $u_{M}\to\infty$ we have
\begin{align*}
	\lim_{\epsilon_{0},u_{M}\to\infty}S^{+}(u_{M}) & = B^x + i B^y-2\kappa S_{0}^{-}+2i\sqrt{2\beta}S_{0}^{z}+2\sum_{j=1}^{L}\left(f_{j}^{-}S_{j}^{x} + if_{j}^{+}S_{j}^{y}\right)
\end{align*}
where
\[
\kappa=\lim_{\epsilon_{0},u_{M}\to\infty}\left[\frac{u_{M}}{\sqrt{\epsilon_{0}}}\right].
\]
By using the expression (\ref{Class I supercharges}) for $A^+$ and noting that
\begin{align}\label{eq: Apm in reparametrisation}
	S^{+}_{k+1}(u) \vert_{\zeta = {+1}} = S_{k}^{+}(u)\vert_{\zeta = {-1}},
\end{align}
the eigenstates become
\begin{align}\label{eq: Eigenstates_diverging_roots}
	\ket{\kappa; u_1,\dots,u_{M-1}} 
	& =\left.2\left[A^{+} - \kappa S_{0}^{+}\right]\ket{-}_0 \otimes \mathcal{S}^{+}(u_{1},\dots,u_{M-1})v_B^{-}\right|_{\zeta =-1}
\end{align}
for $M - 1 = N_s$.
Applying the same limits to the Bethe Ansatz equations (\ref{eq: BAE for spin chain charges}) it follows that the roots satisfy
\begin{align}\label{eq: BAE generic}
	& 2u_{l}+4\beta - i B^x\sqrt{2\beta}+2\sum_{j=1}^{L}\frac{(\epsilon_{j}-\beta)(u_{l}+\beta)}{u_{l}-\epsilon_{j}}s_{j}-2\sum_{m\neq l}^{M-1}\frac{(u_{m}-\beta)(u_{l}+\beta)}{u_{l}-u_{m}}\nonumber\\
	= &\frac{1}{4}\left(\left((M+1)\sqrt{2\beta} - i B^x\right)^2-(B^y)^2\right) \prod_{j=1}^{L}\left(\frac{\epsilon_{j}-u_{l}}{\epsilon_{j}-\beta}\right)^{2s_{j}}\prod_{m\neq l}^{M-1}\frac{u_{m}-\beta}{u_{m}-u_{l}}
\end{align}
and $\kappa$ satisfies
\begin{align*}
	& \kappa^{2}+B^z\kappa+i B^x\sqrt{2\beta}-2\beta-2\sum_{j=1}^{L}(\epsilon_{j}-\beta)s_{j}+2\sum_{m=1}^{M-1}(u_{m}-\beta)\\
	& =-\frac{1}{4}\left(\left((M+1)\sqrt{2\beta}- iB^x\right)^2-(B^y)^2\right)\frac{\prod_{m=1}^{M-1}(u_{m}-\beta)}{\prod_{j=1}^{L}(\epsilon_{j}-\beta)^{2s_{j}}}.
\end{align*}
The energy is
\begin{align}\label{eq: Energy_kappa}
	E & = -B^z/2-\kappa
\end{align}
and the eigenvalues $\{q_i\}$ of the charges $\{Q_i\}$ are
\begin{align*}
	q_i = &-\frac{2\beta}{f_{i}^{+}f_{i}^{-}}s_{i}^{2}+\left(-i B^x\frac{\sqrt{2\beta}}{f_{i}^{+}f_{i}^{-}}+2\frac{f_{i}^{+}}{f_{i}^{-}}\right)s_{i}\\
	&+2\frac{f_{i}^{+}}{f_{i}^{-}}\sum_{m=1}^{M}\frac{u_{m}-\beta}{u_{m}-\epsilon_{i}}s_{i}+2\frac{f_{i}^{+}}{f_{i}^{-}}\sum_{j\neq i}^{L}\frac{\epsilon_{j}-\beta}{\epsilon_{i}-\epsilon_{j}}s_{i}s_{j}.
\end{align*}

\section{The dark states}\label{sec: Dark states} 
There is an alternative way of constructing the eigenstates more analogous to that employed for the $XX$ model in \cite{vcc20}, namely exploiting the connection of the model with a supersymmetric Hamiltonian. This method is more conducive to obtaining the dark states and the conditions admitting their existence and so it will be followed with the construction and characterisation of the dark states. We remark that it is convenient to construct the dark states from both the raising $S^+(u)$ and lowering $S^{-}(u)$ operators. For this reason we will give expressions for both concurrently. We use the notation $\Theta$ to denote either an element of the set $\{+, -\}$ or $\{+1, -1\}$.

\subsection{Construction from supersymmetry}\label{sec: Supersymmetry}
Define the spin raising and lowering operators $S_0^\pm =S_0^x\pm i S_0^y$, and supercharges $\mathcal{A}^{\pm} = S_0^{\mp}A^{\pm}$, where $A^\pm$ is given by (\ref{Class I supercharges}). The supercharges satisfy $(\mathcal{A}^{\pm})^2 = 0$ and $S_0^z \mathcal{A}^{\pm} = - \mathcal{A}^{\pm} S_0^z$. Recall from \cite{vl23} that the Hamiltonian (\ref{H_I}) can be written as
\begin{align} \label{H_I with supercharges}
	H &= B^z S_0^z+\mathcal{A}^++\mathcal{A}^-.
\end{align} 
With this one sees that $H^2$ is related to a supersymmetric Hamiltonian through
\begin{align}\label{H_I squared}
	H^2&= (B^z)^2 I/4+\mathcal{A}^+ \mathcal{A}^- +\mathcal{A}^-\mathcal{A}^+ \\
	&= ((B^x)^2+(B^y)^2+(B^z)^2)I/4+\sum_{j=1}^{L} f_j^+ f_j^- Q_j. \nonumber
\end{align}
Expressed in the eigenbasis of $S^z_0$ yields
\begin{align*}
	H^2 
	&= (B^z)^2/4+(I/2-S^z_0)A^+ A^- +(I/2+S^z_0)A^- A^+ \nonumber\\
	&= \begin{pmatrix}
		 (B^z)^2/4 +A^- A^+ & 0 \cr
		0 & (B^z)^2/4 +A^+ A^-
	\end{pmatrix} \\
	&= \frac{|\vec{B}|^2}{4}I +\begin{pmatrix} \sum_{j=1}^{L} f_j^+ f_j^- \mathcal{Q}^-_j & 0 \cr
		0 & \sum_{j=1}^{L} f_j^+ f_j^- \mathcal{Q}^+_j \end{pmatrix},
\end{align*}
where $\mathcal{Q}_j^{\pm} = \mathcal{Q}_j \vert_{\zeta = \pm1}$. This shows that $H^2$ can be diagonalised with eigenstates of the form
$$\ket{\Psi_{\Theta}} = \ket{\Theta}_0\otimes\ket{\psi_{\Theta}}_B,\quad \Theta \in \{+,-\}$$
where the subscript $0$ denotes a central spin state and $B$ a bath-spin state. Observe that the states $\ket{\psi_\Theta}_B$ are respectively eigenstates of the charges $\mathcal{Q}_j^{-\Theta}$.
Due to the supersymmetry, the eigenstates of $H^2$ are either singlets or doublets. The singlets $\ket{\Psi_{\Theta}}$,
$$\mathcal{A}^+\ket{\Psi_{\Theta}} = \mathcal{A}^-\ket{\Psi_{\Theta}} = 0,$$
are also eigenstates of $H$. Whereas for the doublets $\ket{\Psi_{\Theta}}$,
\begin{align*}
	\mathcal{A}^+\ket{\Psi_{+}} = \Lambda \ket{\Psi_{-}}, \quad
	\mathcal{A}^-\ket{\Psi_{-}} = \Lambda \ket{\Psi_{+}}, \quad \Lambda \in \mathbb R
\end{align*}
we can diagonalise $H$ to obtain the eigenstates 
\begin{align}\label{eq: Bright states from supersymmetry}
	\ket{E_{\pm}} &= \Lambda\ket{\Psi_{+}}+\left(E _{\pm}- B^z/2\right)\ket{\Psi_{-}} \\
	&= \left(\mathcal{A}^-+\left(E_\pm - B^z/2\right)\right)\ket{\Psi_{-}}  \nonumber\\
	&= \left(A^-+\left(E_\pm - B^z/2\right)S_0^-\right)\ket{+}_0\otimes\ket{\psi_{-}}_B \nonumber
\end{align}
where $E_\pm =\pm\sqrt{(B^z)^2/4+\Lambda^2}$. 

From the above construction it is clear that we can obtain the eigenstates of the $XY$ model from those of $\mathcal{Q}_j^{-\Theta}$ or equivalently, due to simultaneous diagonalisability, respectively from $A^{-\Theta}A^\Theta$. 


\subsection{Dark states} The dark states can be constructed from the singlets via the Bethe Ansatz. Doing so one finds these are the states
\begin{align*}
	\left.\ket{\Theta}_0 \otimes \mathcal{S}^{-\Theta}(u_1, \dots, u_M)\, v^\Theta_B\right|_{\zeta = -\Theta}
\end{align*}
in which the $M \in \{0, 1, \dots, N_s\}$ roots satisfy the Bethe Ansatz equations
\begin{align}\label{eq: BAE I}
	2\beta+2\sum_{j=1}^{L}\frac{(u_{l}+\beta)(\epsilon_{j}-\beta)}{u_{l}-\epsilon_{j}}s_{j}-2\sum_{m\neq l}^{M}\frac{(u_{l}+\beta)(u_{m}-\beta)}{u_{l}-u_{m}}=0.
\end{align}
These are eigenstates of $H$ with energy $E = \Theta B^z/2$ if the magnetic field components satisfy
\begin{align}\label{eq: Dark states condition}
	B^x = 0, \quad B^y = (2M - N_s)\sqrt{2\beta}.
\end{align}
 Note that here one requires the solutions for both $\Theta = +1$ and $\Theta = -1$ to obtain a complete set of eigenstates. Recalling the replacement $\zeta \to B^z/\sqrt{\epsilon_0}$ made in the derivation of the XY central spin model, we remark that (\ref{eq: Dark states condition}) is consistent with (\ref{eq: B field cond}).
 
 These states can however be seen as coming from the generic eigenstates of Sec. \ref{sec: Eigenstates} when the magnetic field configuration takes the form (\ref{eq: Dark states condition}).\footnote{See also the appendix \ref{sec: Numerics for dark states} for numerical results confirming this and a heuristic discussion in \ref{sec: Recovering dark states}} To see this note that due to the eigenstates (\ref{eq: Eigenstates_diverging_roots}) being also eigenstates of the charges $\{Q_j\}$ the bath states are eigenstates of $\mathcal{Q}^\pm_j$. Specifically, the bath state of
 \begin{align*}
 	\left.\ket{-}_0 \otimes A^{+}\mathcal{S}^{+}(u_{1},\dots,u_{M-1})v_B^{-}\right|_{\zeta =-1}
 \end{align*}
 is an eigenstate of $\mathcal{Q}^+_j$ and so is equivalent to an eigenstate of the form (\ref{eq: irr Bethe states}) 
and so has Bethe roots being solutions of Bethe Ansatz equations of the form in (\ref{eq: BAE I}).
 
 There is however a simpler way of constructing the dark states which circumvents the Bethe Ansatz. It also allows a clear characterisation of which magnetic field configurations admit dark states.
 
 \subsubsection{Characterisation and construction.}\label{sec: Dark state conditions}
We first show that the eigenstates given in the previous section, namely the singlet states, are the only dark states. As seen in the construction in Sec. \ref{sec: Supersymmetry}, states constructed from singlets are evidently dark states. Examining then the eigenstates obtained from the doublets, namely
 $$\ket{E}=\Lambda\ket{-}_{0}\otimes\ket{\psi_{-}}_{B}+\left(E+B^z/2\right)\ket{+}_{0}\otimes\ket{\psi_{+}}_{B},$$
we see that for these to be dark states requires $(E+B^z/2)\ket{\psi_{+}}_{B}\propto\ket{\psi_{-}}_{B}.$
 If we assume that neither of these is zero this implies that $\ket{\psi_\Theta}_B,\,\Theta \in \{+, -\}$ are eigenstates of both charges $\mathcal{Q}_{j}^{\pm}$ since $\ket{\psi_{\Theta}}_B$ are respectively eigenstates of the charges $\mathcal{Q}_{j}^{ -\Theta}$ due to
 $$A^{+}\mathcal{Q}_{j}^{-}=\mathcal{Q}_{j}^{+}A^{+}.$$
 Hence they are also eigenstates of $\mathcal{Q}_{j}^+-\mathcal{Q}_{j}^- = 2 S_{j}^{z} $ for all $ j=1,\dots,L$. This determines $\ket{\psi_\Theta}_B,\,\Theta \in \{+,-\}$ and would lead to a contradiction if we took e.g. $B^x\gg1$ since this holds for all $B^x$. This also means that dark states can only occur if $\Lambda\ket{\psi_{-}}_{B}=0$ or $(E+B^z/2)\ket{\psi_{+}}_{B}=0$ in which case the central spin state is respectively $\ket{\Theta}_{0}$. Furthermore, by the supersymmetry this means that $\ket{\Theta}_{0}\otimes\ket{\psi_{\Theta}}_{B}$ is a singlet (as these were originally doublets). 
 
 Next we construct the dark states. From the above argument these are the singlet states
 $$\ket{\Phi_{\Theta}} = \ket{\Theta}_0\otimes\ket{\phi_{\Theta}}_B,\quad \Theta \in \{+,-\}$$
with
 $$A^\Theta\ket{\phi_{\Theta}} = 0.$$ Set $\mathfrak{f}_j^{\pm}:=(f_j^{-}\pm f_j^{+})/2$, so that we can write the local operators of $A^{\pm}$ in terms of $S_j^\pm$ as
 \begin{align*}
 	f^{-}_j S_j^{x}\pm i f_j^{+}S_j^{y} & =\mathfrak{f}_j^{\pm}S_j^{+}+\mathfrak{f}_j^{\mp}S_j^{-}.
 \end{align*}
 We define operators
 \begin{align*}
 	E_j^\pm & =-\frac{\sqrt{\mathfrak{f}_j^{\pm}}}{2\sqrt{\mathfrak{f}_j^{\mp}}}S_j^{+}+S_j^{z}+\frac{\sqrt{\mathfrak{f}_j^{\mp}}}{2\sqrt{\mathfrak{f}_j^{\pm}}}S_j^{-}, \;
 	H_j^\pm =\frac{\sqrt{\mathfrak{f}_j^{\pm}}}{\sqrt{\mathfrak{f}_j^{\mp}}}S_j^{+}+\frac{\sqrt{\mathfrak{f}_j^{\mp}}}{\sqrt{\mathfrak{f}_j^{\pm}}}S_j^{-},\\
 	F_j^\pm  &=\frac{\sqrt{\mathfrak{f}_j^{\pm}}}{2\sqrt{\mathfrak{f}_j^{\mp}}}S_j^{+}+S_j^{z}-\frac{\sqrt{\mathfrak{f}_j^{\mp}}}{2\sqrt{\mathfrak{f}_j^{\pm}}}S_j^{-},
 \end{align*}
that satisfy the $\mathfrak{sl}(2)$ relations
 \begin{align*}
 	[H_j^\pm ,E_j^\pm ]=2E_j^\pm , \quad [E_j^\pm ,F_j^\pm]=H_j^\pm , \quad [H_j^\pm ,F_j^\pm ]=-2F_j^\pm .
 \end{align*}
 Noting that
 \begin{align*}
 	A^{\pm} &=(B^x \pm i B^y)I/2+\sum_{j=1}^{L}\sqrt{\mathfrak{f}^{+}_j \mathfrak{f}^{-}_j}\,H_{j}^\pm \\
 	&= (B^x \pm i B^y)I/2+i\sum_{j=1}^{L}\sqrt{\beta/2}\,H_{j}^\pm
 \end{align*}
 we can respectively diagonalise $A^\pm$ with the respective eigenvectors
 \begin{align*}
 	&\ket{n_{1},\dots,n_{L}}^{+} = \prod_{j=1}^{L}(E_{j}^+)^{n_{j}}\chi^+,\\ &\ket{n_{1},\dots,n_{L}}^{-} = \prod_{j=1}^{L}(F_{j}^-)^{n_{j}}\chi^-,
 \end{align*}
 where the respectively lowest and highest weight vector is defined as
 \begin{align*}
 	&\chi^{\pm}=\chi_{1}^\pm\otimes\cdots \chi_{L}^\pm, \\
 	&F_{j}^+\chi_{j}^+=0,\; H_{j}^+ \chi_{j}^+ = -2s_{j}\chi_{j}^+ ;\; -2s_{j}\leq n_{j}\leq2s_{j}, \\
 	&E_{j}^-\chi_{j}^-=0,\; H_{j}^- \chi_{j}^- = 2s_{j}\chi_{j}^- ;\; -2s_{j}\leq n_{j}\leq2s_{j}.
 \end{align*}
It is found that $A^\pm$ has respective eigenvalues
 \begin{align*}
 	(B^x \pm i B^y)/2\pm\left(-i\sqrt{2\beta}\sum_{j=1}^{L}n_{j}+\frac{N_{s}}{2}i\sqrt{2\beta}\right),\quad -2s_{j}\leq n_{j}\leq2s_{j}.
 \end{align*}
 We see that these are zero, i.e. the corresponding states $$\ket{\Phi_{\Theta}} = \ket{\Theta}_0 \otimes \ket{n_{1},\dots,n_{L}}^{\Theta}$$ are dark states, when
 \begin{align*}
 	\frac{-i B^x + \Theta B^y}{\sqrt{2\beta}} & = \Theta\left(2\sum_{j=1}^{L}n_{j}-N_s\right)\in\{-N_{s}/2,\dots,N_{s}/2\}
 \end{align*}
 holds. Due to the Hamiltonian being Hermitian this requires $B^x = 0$ and we obtain (\ref{eq: Dark states condition}) after making the identification $M = \sum_{j=1}^{L}n_{j}$.
 
 We briefly note that for the $XX$ model case the dark states are zero eigenvalue eigenstates of
 $$A^{\pm}=(B^{x}\pm iB^{y})I/2+\sum_{j=1}^{L}\sqrt{\epsilon_{j}}S_{j}^{\pm}.$$
 Hence, it follows that $A^{\pm}$ is strictly upper or lower triangular if and only if $B^{x}=B^{y}=0$, and so dark states can only occur if $B^{x}=B^{y}=0$, namely that treated in \cite{vcc20}.
 
\subsubsection{Quadratic relations.} We mention that some of the roots becoming equal to $\beta$ is consistent with the quadratic relations of the charges $\mathcal{Q}_j$ as seen in \cite{vl23}. Restricting to spin-$1/2$ for all of the spins, the charges of the XYZ Richardson-Gaudin model $\mathcal{Q}_j$\footnote{With a constant shift introduced to make the relations cleaner.}
\begin{align*}
	\mathcal{Q}_{j} & = \zeta S_{j}^{z}+\frac{B^x}{f_{j}^{+}}S_{j}^{x}+\frac{B^y}{f_{j}^{-}}S_{j}^{y}\\
	& \quad+2\sum_{k\neq j}^{L}\frac{1}{\epsilon_{j}-\epsilon_{k}}\left(f_{j}^{+}f_{k}^{-}S_{j}^{x}S_{k}^{x}+f_{j}^{-}f_{k}^{+}S_{j}^{y}S_{k}^{y}+f_{k}^{+}f_{k}^{-}\left(S_{j}^{z}S_{k}^{z}-\frac{1}{4}\right)\right)
\end{align*}
satisfy the following quadratic relations
\begin{align}\label{eq: quadratic relations}
	\mathcal{Q}_{j}^{2}= & \frac{1}{4}\left(\zeta^{2}+\frac{(B^x)^{2}}{(f^+_j)^{2}}+\frac{(B^y)^{2}}{(f^-_j)^{2}}\right)-\sum_{k\neq j}^L\frac{f^+_j f^-_j}{\epsilon_{j}-\epsilon_{k}}\left(\mathcal{Q}_{j}-\mathcal{Q}_{k}\right)\nonumber\\
	&\quad+\frac{1}{4}\sum_{k\neq j}^L\frac{(f^+_j f^-_k-f^-_j f^+_k)^{2}}{(\epsilon_{j}-\epsilon_{k})^{2}}
\end{align}
with these also holding true for their eigenvalues
\begin{align*}
	\mathfrak{q}_{j}^{2}= & \frac{1}{4}\left(\zeta^{2}+\frac{(B^x)^{2}}{(f^+_j)^{2}}+\frac{(B^y)^{2}}{(f^-_j)^{2}}\right)-\sum_{k\neq j}^L\frac{f^+_j f^-_j}{\epsilon_{j}-\epsilon_{k}}\left(\mathfrak{q}_{j}-\mathfrak{q}_{k}\right)\\
	&\quad+\frac{1}{4}\sum_{k\neq j}^L\frac{(f^+_j f^-_k-f^-_j f^+_k)^{2}}{(\epsilon_{j}-\epsilon_{k})^{2}}.
\end{align*}
With a slightly more general ansatz than that of \cite{vl23}, for a parametrisation of the eigenvalues in terms of the parameters $\{u_m\}_{m=1}^M$
\begin{align}\label{eq: eigenvalue param}
	\mathfrak{q}_{j} & =C\frac{f_{j}^{+}}{f_{j}^{-}}+D\frac{f_{j}^{-}}{f_{j}^{+}}-\sum_{m=1}^{M}\frac{f_{j}^{+}f_{j}^{-}}{\epsilon_{j}-u_{m}}+\frac{1}{2}\sum_{k\neq j}^{L}\frac{f_{j}^{+}f_{j}^{-}-f_{k}^{+}f_{k}^{-}}{\epsilon_{j}-\epsilon_{k}}\\
	& =\alpha_{j}(\epsilon_{j})\frac{\epsilon_{j}}{f_{j}^{+}f_{j}^{-}}-\sum_{m=1}^{M}\frac{f_{j}^{+}f_{j}^{-}}{\epsilon_{j}-u_{m}}+\frac{1}{2}\sum_{k\neq j}^{L}\frac{f_{j}^{+}f_{j}^{-}-f_{k}^{+}f_{k}^{-}}{\epsilon_{j}-\epsilon_{k}} \nonumber
\end{align}
where
\begin{align*}
	\alpha(\epsilon_{j}) & =\alpha_{0}+\frac{\alpha_{1}}{\epsilon_{j}} =(C+D)+(C-D)\frac{\beta}{\epsilon_{j}},
\end{align*}
the quadratic relations give consistency equations for the parameters $\{u_m\}_{m=1}^M$. These are the Bethe Ansatz equations (\ref{eq: BAE for spin chain charges}) for
\begin{align*}
	C = \frac{1}{2} \left(2M-L+1-\frac{i  B^x}{\sqrt{2\beta }}- \zeta\right),\quad D = \frac{i B^x}{2\sqrt{2\beta }}.
\end{align*}
From this we see that when some of the $M = L$ roots become equal to $\beta$ the eigenvalues and Bethe Ansatz equations reduce to (\ref{eq: standard BAE for spin chain charges}) when the magnetic field satisfies (\ref{eq: B field cond}).

\section{Conclusion}\label{sec: Conclusion}

We have derived the $XY$ central spin model Hamiltonian $H_{XY}$ (\ref{H_I}) and charges given in \cite{vl23}, as a limit of the $XYZ$ Richardson-Gaudin model charges with self-interaction $\mathcal{Q}_{i}$ obtained by Skrypnyk \cite{s22}. This allowed the diagonalisation results found in  \cite{s22} for the charges $\mathcal{Q}_{i}$ via a modified Algebraic Bethe Ansatz to be used to diagonalise $H_{XY}$. However, in order for these results to connect with those found for the $XX$ model we regularised the charges $\mathcal{Q}_{i}$ via a reparametrisation. Lastly, we showed that the dark states of the $XX$ model without an in-plane magnetic field \cite{df22, vcc20} can be seen as a special case of dark states of the $XY$ model for special configurations of the magnetic field. 


These results open the way for finding form factors and correlation functions for the $XY$ central spin model as well as probing the dynamics for large systems. The characterisation of the magnetic field configurations supporting dark states and their emergence via special limits of the parameter $\kappa$ can provide experimental criteria for preparing these states with high-fidelity and determining how far a bright state is from being a product state.

Another extension of this work would be to the case of a spin-1 central spin. Here it has been shown that the $XX$ central spin model is integrable for an out-of-plane magnetic field $\cite{tlpcc23}$ and more recently for an arbitrarily oriented magnetic field $\cite{nmf24}$. Integrability for the $XY$ model can also be shown through a minor modification of the charges found in $\cite{nmf24}$. In the latter work it was numerically demonstrated that dark states can emerge at asymptotically large coupling, just as seen in the central spin-1/2 model for an arbitrarily oriented magnetic field $\cite{df22}$. This raises the possibility of the existence of these states for certain special magnetic field configurations of the $XX$ and $XY$ spin-1 central spin models. However, determining whether these dark states actually exist is complicated by the fact that the relationship between the model and the known class of Richardson-Gaudin models is not yet understood.

\section*{Acknowledgments}
The authors thank Pieter Claeys and Taras Skrypnyk for helpful comments that enabled progress on this research problem. The authors also acknowledge the traditional owners of
the land on which The University of Queensland at St. Lucia operates,
the Turrbal and Jagera people. This work was supported by the Australian Research Council through Discovery Project DP200101339.

 \section{Appendix}\label{sec: Appendix}
 
 \subsection{Reparametrisation.}\label{sec: Reparametrised operators} 
 In the derivation of the expressions for the eigenstates in Sec. \ref{sec: Eigenstates} from those of \cite{s22} we employed a reparametrisation of the model parameters.  Write the charges $\{\mathcal{Q}_j\}$ (\ref{QQ_i}) in terms of the parameters $\{\phi_{j}\}$ and magnetic field components $\{\tilde{B^x}, \tilde{\zeta}, \tilde{B^z}\}$
 \begin{align*}
 	\tilde{\mathcal{Q}_{j}} & =\tilde{B^z}S_{j}^{z}+\frac{\tilde{B^x}}{h_{j}^{+}}S_{j}^{x}+\frac{\tilde{\zeta}}{h_{j}^{-}}S_{j}^{y}+\frac{h_{j}^{-}}{h_{j}^{+}}S_{j}^{x}S_{j}^{x}+\frac{h_{j}^{+}}{h_{j}^{-}}S_{j}^{y}S_{j}^{y}\\
 	& \quad+2\sum_{k\neq j}^{L}\frac{1}{\phi_{j}-\phi_{k}}(h_{j}^{+}h_{k}^{-}S_{j}^{x}S_{k}^{x}+h_{j}^{-}h_{k}^{+}S_{j}^{y}S_{k}^{y}+h_{k}^{+}h_{k}^{-}S_{j}^{z}S_{k}^{z}),
 \end{align*}
 where $h^\pm_j = \sqrt{\phi_j \pm \beta}$. Then the reparametrisation that was applied is the following
 \begin{align}\label{eq: Reparametrisation}
 	&\epsilon_j = \beta + \frac{4\beta^2}{\phi_j - \beta}, \nonumber\\
 	&\tilde{B^x}=-B^x,\quad\tilde{\zeta}=-\zeta\sqrt{2\beta},\quad\tilde{B^z}=B^y/\sqrt{2\beta}.
 \end{align}
 The spectral parameter $u$ in the argument of the Lax generators was similarly reparametrised as
 $$\quad u = \beta + \frac{4\beta^2}{w - \beta},$$
 where the Lax generators in \cite{s22} correspond to those of Sec. \ref{Lax algebra relations} via
 \begin{align*}
 	S^{z}(w) = 2\beta S^{-}(u(w)), \; S^{+}(w) = \frac{1}{\sqrt{2\beta}}S^{+}(u(w)), \; S^{-}(w)  = \frac{1}{\sqrt{2\beta}}S^{-}(u(w)),
 \end{align*}
and these are identified with $\{A(w), B(w), C(w), D(w)\}$ as
 \begin{align*}
 	S^{z}(w) = \frac{A(w)-D(w)}{2}, \quad S^{+}(w) = C(w), \quad S^{-}(w) = D(w).
 \end{align*}
 
 The reparametrisation is required to regularise the eigenstates in the isotropic limit $\beta \to 0$ in order to recover the eigenstates of the $XX$ model as seen in \cite{vcc20}. It is enabled by the observation that under this reparametrisation the charges (\ref{QQ_i}) take the form (after making the rotation $S^y\to S^z,\; S^z\to-S^y$ and adding $\mathbf{S}_i^2$)
 \begin{align}\label{eq: Map between two parameterisations of the charges}
 	\tilde{\mathcal{Q}}_{i} & =-\frac{f_{i}^{-}}{\sqrt{2\beta}}\bigg\{-\frac{f_{i}^{+}}{f_{i}^{-}}s_{i}(s_{i}+1)-\frac{\tilde{\zeta}}{\sqrt{2\beta}} S^z_i-\frac{\tilde{B^x}}{f^+_i}S^x_i+\tilde{B^z}\frac{\sqrt{2\beta}}{f^-_i}S^y_i +\frac{f_{i}^{-}}{f_{i}^{+}}S_{i}^{x}S_{i}^{x} \nonumber\\
 	&\quad+\frac{f_{i}^{+}}{f_{i}^{-}}S_{i}^{y}S_{i}^{y}+2\sum_{j\neq i}^{L}\frac{1}{\epsilon_{i}-\epsilon_{j}}\left(f^+_i f^-_j S^x_i S^x_j+f^-_i f^+_j S^y_i S^y_j+f^+_j f^-_jS^z_i S^z_j\right)\bigg\} \nonumber\\
 	& =-\frac{f_{i}^{-}}{\sqrt{2\beta}}\left(\mathcal{Q}_{i}-\frac{f_{i}^{+}}{f_{i}^{-}}s_{i}(s_{i}+1)\right)
 \end{align}
 where $\{\mathcal{Q}_{i}\}$ are simply the charges (\ref{QQ_i}).
 
 \subsection{Eigenstates of the $XX$ central spin model}\label{sec: Eigenstates of XX model}
 
Due to the reparametrised expressions for the eigenstates and eigenvalues of the $XYZ$ Richardson-Gaudin model charges all being non-singular in the limit $\beta \to 0$, we can rederive the diagonalisation results for the $XX$ model. This will also give us an idea of how the dark states emerge in the limit of the special magnetic field configurations (\ref{eq: Dark states condition}) of Sect. \ref{sec: Dark states}.
 
 Taking the limit as $\beta \to 0$, the eigenstates of the charges
 \begin{align*}
 	\mathcal{Q}_{i} = \zeta S^z_i+\frac{B^x}{\sqrt{\epsilon_i}} S^x_i+\frac{B^y}{\sqrt{\epsilon_i}}S^y_i +S_{i}^{x}S_{i}^{x}+S_{i}^{y}S_{i}^{y}\\
 	\qquad+2\sum_{j\neq i}^{L}\frac{1}{\epsilon_{i}-\epsilon_{j}}\left(\sqrt{\epsilon_i}\sqrt{\epsilon_j} S^x_i S^x_j+\sqrt{\epsilon_i}\sqrt{\epsilon_j} S^y_i S^y_j+\epsilon_j S^z_i S^z_j\right)
 \end{align*}
 become
 \begin{align*}
 	\ket{u_1,\dots,u_M}_{-\Theta} & = S^{-\Theta}(u_{1})S^{-\Theta}(u_{2})\cdots S^{-\Theta}(u_{M})v^{\Theta},
 \end{align*}
 with $\Theta \in \{+,-\}$.
 Here
 \begin{align*}
 	S^{-\Theta}(u) & = B^x-\Theta i B^y +2\sum_{j=1}^{L}\frac{u\sqrt{\epsilon_{j}}}{u-\epsilon_{j}}S_j^{-\Theta}
 \end{align*}
 and the reference states $v^{\Theta} =v_{0}^{\Theta}\otimes v_{1}^{\Theta}\otimes\cdots\otimes v_{L}^{\Theta}$ now satisfy
 \begin{align*}
 	S_{k}^{z}v_{k}^+= s_{k}v_{k},\; S_{k}^{+}v_{k}^{+}=0;\quad
 	S_{k}^{z}v_{k}^- = -s_{k}v_{k},\; S_{k}^{-}v_{k}^{-}=0,
 \end{align*}
 or in other words $$v^{\Theta} = \ket{\Theta s_{1},\dots,\Theta s_{L}}.$$ The Bethe roots in turn now satisfy
 \begin{align}\label{eq: BAE for XXZ model}
 	1+\Theta \zeta +&2\sum_{j=1}^{L}\frac{\epsilon_{j}}{u_{l}-\epsilon_{j}}s_{j}-2\sum_{m\neq l}^{M}\frac{u_{m}}{u_{l}-u_{m}} \nonumber\\
 	 &= -\frac{(B^x)^2+(B^y)^2}{4}\frac{\prod_{j= 1}^{L}\left(u_{l}^{-1}-\epsilon_{j}^{-1}\right)^{2s_{j}}}{\prod_{m\neq l}^{M}(u_{l}^{-1}-u_{m}^{-1})}
 \end{align}
 with $M = \sum_{j=1}^L 2s_j$, while the eigenvalues $\{\mathfrak{q}^{-\Theta}_j\}$ of the charges $\{\mathcal{Q}_j\}$ are
 \begin{align*}
 	\mathfrak{q}^{-\Theta}_{i} & = \left(1+\Theta  \zeta \right)s_{i}+2\sum_{m=1}^{M}\frac{u_{m}}{u_{m}-\epsilon_{i}}s_{i}+2\sum_{j\neq i}^{L}\frac{\epsilon_{j}}{\epsilon_{i}-\epsilon_{j}}s_{i}s_{j}.
 \end{align*}
 
 To obtain the eigenstates of the $XX$ central spin model we can simply use the results for the $XY$ model in the limit $\beta \to 0$. For a magnetic field with a non-zero in-plane component the eigenstates are entangled states of the central spin and bath, the bright states of \cite{vcc20}, taking the form
 \begin{align*}
 	\ket{\kappa; u_1,\dots,u_{M-1}}_{\Theta} = &\left[B^x-\Theta i B^y +2\sum_{j=1}^{L}\sqrt{\epsilon_{j}}S_{j}^{-\Theta}-2\kappa S_{0}^{-\Theta}\right]\ket{\Theta}_0\\
 	&\otimes\left.\mathcal{S}^{-\Theta}(u_{1},u_{2},\dots,u_{M-1})\ket{\Theta s_{1},\dots,\Theta s_{L}}\right|_{\zeta =\Theta}
 \end{align*}
 with the $M-1 = \sum_{j=1}^L 2s_j$ roots satisfying the Bethe Ansatz equations
 \begin{align*}
 	& 2+2\sum_{j=1}^{L}\frac{\epsilon_{j}}{u_{l}-\epsilon_{j}}s_{j}-2\sum_{m\neq l}^{M-1}\frac{u_{m}}{u_{l}-u_{m}}
 	& =-\frac{(B^x)^2+(B^y)^2}{4}\frac{\prod_{j=1}^{L}(u_{l}^{-1}-\epsilon_{j}^{-1})^{2s_{j}}}{\prod_{m\neq l}^{M-1}(u_{l}^{-1}-u_{m}^{-1})}
 \end{align*}
 and the equation for $\kappa$ coming from the Bethe Ansatz equation for $u_M$,
 \begin{align}\label{eq: kappa XX}
 	\kappa^{2}-\Theta B^z \kappa-2\sum_{j=1}^{L}\epsilon_{j}s_{j}+2\sum_{m=1}^{M-1}u_{m} & =-\frac{(B^x)^2+(B^y)^2}{4}\frac{\prod_{m=1}^{M-1}u_{m}}{\prod_{j=1}^{L}(\epsilon_{j})^{2s_{j}}}.
 \end{align}
 The energy is
 \begin{align*}
 	E & =\Theta B^z/2-\kappa
 \end{align*}
 while the eigenvalues $\{q_i\}$ of the charges $\{Q_i\}$ are
 \begin{align*}
 	q_i & = 2s_{i}-2\sum_{m=1}^{M-1}\frac{u_{m}}{\epsilon_{i}-u_{m}}s_{i}+2\sum_{j\neq i}^{L}\frac{\epsilon_{j}}{\epsilon_{i}-\epsilon_{j}}s_{i}s_{j}.
 \end{align*}
 Observe here that if $\kappa$  is a solution to the first equation of $(\ref{eq: kappa XX})$ then $\kappa'=\kappa-B^z$ is a solution to the second (respectively, if $\kappa$  is a solution to the second equation of $(\ref{eq: kappa XX})$ then $\kappa'=\kappa+B^z$ is a solution to the first equation), due to the Bethe equations for the roots being the same for both. These have identical energy and charge eigenvalues. Therefore, based on the argument of the correspondence between eigenstates constructed using raising or lowering operators, they are proportional to the same eigenstate.
 We note these expressions match those of \cite{cdv16} and \cite{vcc20}, where in the former $s_j = 1/2$ and in the latter $B^x = B^y = 0$ .
 
 \subsubsection{Recovering the dark states for $B^x = B^y = 0$.}\label{sec: Recovering dark states}
 
 As argued at the end of Sect. \ref{sec: Dark state conditions}, in the case of a non-zero in-plane magnetic field, $(B^x)^2+(B^y)^2 > 0$, there are no dark states. We find numerically (see also Sect. \ref{sec: Numerics for dark states}) that the dark states of the $XX$ model with no in-plane field are recovered in the limit as $(B^x)^2+(B^y)^2 \to 0$ via $\kappa \to \Theta  B^z$ or $\kappa \to 0$ for some of the eigenstates. For simplicity we will only discuss the case where for all the bath spins $s_j = 1/2$.
 
 In the case where $\kappa \to \Theta  B^z$, the states become
 $$\ket{\psi_{\Theta}}	=2\left[\sum_{j=1}^{L}\sqrt{\epsilon_{j}}S_{j}^{-\Theta} + B^z S_{0}^{-\Theta}\right]\ket{\Theta}_0\otimes \left.\mathcal{S}^{-\Theta}(u_{1},u_{2},\dots,u_{\mathfrak{M}-1})\ket{\Theta s_{1},\dots,\Theta s_{L}}\right|_{\zeta=\Theta}$$
 where, from (\ref{eq: BAE for XXZ model}), the roots satisfy
 \begin{align*}
 	\underbrace{1+\Theta \left(\Theta \right)}_{=2}+2\sum_{j=1}^{L}\frac{\epsilon_{j}}{u_{l}-\epsilon_{j}}s_{j}-2\sum_{m\neq l}^{\mathfrak{M}-1}\frac{u_{m}}{u_{l}-u_{m}}  = 0,
 \end{align*}
 with $\mathfrak{M}\leq M$ due to some of the roots becoming zero and their corresponding raising and lowering operators being asymptotically $S^{-\Theta}(u)\propto c(B^x,B^y)I$. This follows by noting that the Bethe states will become eigenstates of the total $S^z$ operator, as the raising and lowering operators are proportional to products of total spin raising or lowering operators in the zero in-plane field limit, and since the $XX$ model in this limit has a global $U(1)$ symmetry. 
 
 Using the correspondence coming from the symmetry of the charges, $\{S^z \to -S^z,\; i \to -i,\; \zeta \to -\zeta\}$, this can also be written as
 $$\ket{\psi_{\Theta}}	\propto 2\left[\sum_{j=1}^{L}\sqrt{\epsilon_{j}}S_{j}^{-\Theta}+ B^z S_{0}^{-\Theta}\right]\ket{\Theta}_0\otimes \left.\mathcal{S}^{\Theta}(\tilde{u}_{1},\tilde{u}_{2},\dots,\tilde{u}_{\mathcal{M}-1})\ket{-\Theta s_{1},\dots,-\Theta s_{L}}\right|_{\zeta=-\Theta}$$
 with the roots $\{\tilde{u}_m\}_{m=1}^{\mathcal M-1}$ satisfying
 \begin{align*}
 	\underbrace{1+\Theta \left(-\Theta\right)}_{=0}+2\sum_{j=1}^{L}\frac{\epsilon_{j}}{\tilde{u}_{l}-\epsilon_{j}}s_{j}-2\sum_{m\neq l}^{\mathcal{M}-1}\frac{\tilde{u}_{m}}{\tilde{u}_{l}-\tilde{u}_{m}}  = 0.
 \end{align*}
 Note that the bath component of the state now satisfies (see for example \cite{vcc20})
 \begin{align*}
 	\underbrace{\left(\sum_{j=1}^{L}\sqrt{\epsilon_{j}}S_{j}^{-\Theta}\right)}_{A^{-\Theta}}\left.\mathcal{S}^{\Theta}(\tilde{u}_{1},\tilde{u}_{2},\dots,\tilde{u}_{\mathcal{M}-1})\ket{-\Theta s_{1},\dots, -\Theta s_{L}}\right|_{\zeta=-\Theta} = 0,
 \end{align*}
 giving the dark state
 \begin{align*}
 	\ket{\psi_{\Theta}} &\propto \ket{-\Theta}_0\otimes \left.\mathcal{S}^{\Theta}(\tilde{u}_{1},\tilde{u}_{2},\dots,\tilde{u}_{\mathcal{M}-1})\ket{-\Theta s_{1},\dots, -\Theta s_{L}}\right|_{\zeta=-\Theta}.
 \end{align*}
 
 As explained in Sect. \ref{sec: Eigenstates of XX model}, the states with $\kappa' = \kappa +  B^z$ constructed with lowering operators correspond to states with $\kappa$ constructed with raising operators (and vice versa for $\kappa' = \kappa -  B^z$ and $\kappa$). So the states constructed with raising operators for which $\kappa \to 0$ are the same dark states as those constructed with lowering operators with the behaviour $\kappa \to + B^z$ (and vice versa for $\kappa \to 0$ and $\kappa \to - B^z$) that we just found. However, for completeness we will sketch here the behaviour in this limit. For conciseness let $\rho_{-\Theta} =(B^x-\Theta iB^y)/2$  so that the state asymptotically becomes 
 \begin{align*}
 	\ket{\psi_{\Theta}}	\sim\ket{\Theta}_0\otimes&\left(\rho_{-\Theta}+\sum_{j=1}^{L}\sqrt{\epsilon_{j}}S_{j}^{-\Theta}\right)\left(\rho_{-\Theta}+\sum_{j=1}^{L}\frac{u_{1}\sqrt{\epsilon_{j}}}{u_{1}-\epsilon_{j}}S_{j}^{\Theta}\right)\times \\
 	&\cdots\left(\rho_{-\Theta}+\sum_{j=1}^{L}\frac{u_{\mathfrak{M}-1}\sqrt{\epsilon_{j}}}{u_{\mathfrak{M}-1}-\epsilon_{j}}S_{j}^{\Theta}\right)\ket{\Theta s_{1},\dots,\Theta s_{L}}
 \end{align*}
 where again some of the roots $\{u_{i}\}_{i=1}^{M-1}$ go to zero. Expanding this out, the state has the form (for $s_{i}=1/2$) 
 \begin{align*}
 	\ket{\psi_{\Theta}}	\sim\ket{\Theta}_0\otimes&\left(\rho_{-\Theta}^{\mathfrak{M}}+\rho_{-\Theta}^{\mathfrak{M}-1}\sum_{i_{1}}c_{i_{1}}S_{i_{1}}^{\mp}+\cdots+\rho_{-\Theta}^{\mathfrak{M}-K}\sum_{i_{1}<\cdots<i_{K}}c_{i_{1}\cdots i_{K}}S_{i_{1}}^{-\Theta}\cdots S_{i_{K}}^{-\Theta}\right. \\
 	&\left.+\cdots+\sum_{i_{1}<\cdots<i_{\mathfrak{M}}}^{L}c_{i_{1}\cdots i_{\mathfrak{M}}}S_{i_{1}}^{-\Theta}\cdots S_{i_{\mathfrak{M}}}^{-\Theta}\right)\ket{\Theta s_{1},\dots,\Theta s_{L}}
 \end{align*}
 where $$c_{i_{1}\cdots i_{K}}=\sum_{j_{1}<\cdots<j_{K-1}}\sum_{\sigma\in S_{K}}\epsilon_{i_{\sigma(1)}}\frac{u_{j_{1}}\sqrt{\epsilon_{i_{\sigma(2)}}}}{u_{j_{1}}-\epsilon_{i_{\sigma(2)}}}\cdots\frac{u_{j_{K-1}}\sqrt{\epsilon_{i_{\sigma(K)}}}}{u_{j_{K-1}}-\epsilon_{i_{\sigma(K)}}}.$$
 For a dark state with $\mathcal{M}$ roots, all the $c_{i_{1}\cdots i_{K}}$ for $K>\mathcal{M}$ are zero while the leading order coefficients are
 $$c_{i_{1}\cdots i_{\mathcal{M}}}	=\sum_{\sigma\in S_{\mathcal{M}}}\left(\frac{\tilde{u}_{1}\sqrt{\epsilon_{i_{\sigma(1)}}}}{\tilde{u}_{1}-\epsilon_{i_{\sigma(1)}}}\cdots\frac{\tilde{u}_{\mathcal{M}}\sqrt{\epsilon_{i_{\sigma(\mathcal{M})}}}}{\tilde{u}_{\mathcal{M}}-\epsilon_{i_{\sigma(\mathcal{M})}}}\right).$$
 The state in the limit then becomes equal to the dark state
 \begin{align*}
 	\ket{\psi_{\Theta}} \propto \ket{\Theta}_0\otimes\left(\sum_{j=1}^{L}\frac{\tilde{u}_{1}\sqrt{\epsilon_{j}}}{\tilde{u}_{1}-\epsilon_{j}}S_{j}^{\Theta}\right)\cdots\left(\sum_{j=1}^{L}\frac{\tilde{u}_{\mathcal{M}}\sqrt{\epsilon_{j}}}{\tilde{u}_{\mathcal{M}}-\epsilon_{j}}S_{j}^{\Theta}\right)\ket{\Theta s_{1},\dots,\Theta s_{L}}
 \end{align*}
 with $\mathcal{M} \leq \mathfrak{M}$ and where the roots satisfy $$2\sum_{j=1}^{L}\frac{\epsilon_{j}}{\tilde{u}_{l}-\epsilon_{j}}s_{j}-2\sum_{m\neq l}^{\mathcal{M}}\frac{\tilde{u}_{m}}{\tilde{u}_{l}-\tilde{u}_{m}}	=0.$$
 
 \subsection{Emergence of dark states}\label{sec: Numerics for dark states}

In the following we show numerical examples of dark states emerging due to $\kappa \to 0, \Theta  B^z$ when condition (\ref{eq: Dark states condition}) is met. We set the model parameters as $L = 3, \beta = 0.1, B^z = 1$, and\footnote{This $\underline{\epsilon}$ was generated in Mathematica with SeedRandom 1324 using RandomReal in the range 0.4 to 1.3.}
\begin{align*}
	\underline{\epsilon} = \{0.5869853530170386, 1.270553831408777, 1.2426482643150194\},
\end{align*}
and only look at the case where all spins are $s_i = 1/2$. Then for different $B^y$ we change $B^x \to 0$ so that we approach conditions where dark states emerge. As we use states constructed from $v^+$ with the lowering operators $S^-$, we should observe for dark states that $\kappa \to 0, +B^z $. We label states by the eigenvalue $q_1$ of $Q_1$.

\subsubsection{Dark states for $B^y = L\sqrt{2\beta}$.}

In this case there are only two dark states in the limit, both with the same eigenvalues for the charges but with different energies.
\begin{center}
	\begin{tabular}{||c | c | c | c||} 
		\hline
		$B^x/2$ & $\kappa_1$ & $\kappa_2$ & $q_1$ \\ [0.5ex] 
		\hline\hline
		$10^{0}$ & $-5.72938\times 10^{-2}$ & $1+5.72938 \times 10^{-2}$ & $-2.57221$\\
		\hline
		$10^{-1}$ & $-6.70952\times 10^{-5}$ & 
		$1 + 6.70952 \times 10^{-5}$ & $-2.14310$\\ 
		\hline
		$10^{-2}$  & $-6.33204 \times 10^{-7}$& $1 + 6.33204 \times 10^{-7}$ & $-2.13838$\\ 
		\hline
		$10^{-3}$  & $-6.32830 \times 10^{-9}$ & $1 + 6.32830 \times 10^{-9}$ & $-2.13833$\\ 
		\hline
		$10^{-4}$  & $-6.32829 \times 10^{-11}$ & $1 + 6.32829 \times 10^{-11}$ & $-2.13833$ \\ [1ex] 
		\hline
	\end{tabular}
\end{center}

\subsubsection{Dark states for $B^y = \left(L-2\right) \sqrt{2\beta}$.}

In this case there are six dark states in the limit.
\begin{center}
	\begin{tabular}{||c | c | c | c||} 
		\hline
		$B^x/2$ & $\kappa_1$ & $\kappa_2$ &  $q_1$ \\ [0.5ex] 
		\hline\hline
		$10^{-1}$ & $-1.03242\times 10^{-5}$ & 
		$1 + 1.03242 \times 10^{-5}$ & $-1.90365$\\ 
		\hline
		$10^{-2}$  & $-9.33203 \times 10^{-8}$& $1 + 9.33203 \times 10^{-8}$ & $-1.89852$ \\ 
		\hline
		$10^{-3}$  & $-9.32225 \times 10^{-10}$ & $1 + 9.32225 \times 10^{-10}$ & $-1.89846$\\ 
		\hline
		$10^{-4}$  & $-9.30916 \times 10^{-12}$ & $1 + 9.30916 \times 10^{-12}$ & $-1.89846$\\ [1ex] 
		\hline
	\end{tabular}
\end{center}

\begin{center}
	\begin{tabular}{||c | c | c | c||} 
		\hline
		$B^x/2$ & $\kappa_1$ & $\kappa_2$ &  $q_1$ \\ [0.5ex] 
		\hline\hline
		$10^{-1}$ & $-2.97399\times 10^{-3}$ & 
		$1 + 2.97399 \times 10^{-5}$ & $-9.88279 \times 10^{-2}$\\ 
		\hline
		$10^{-2}$  & $-2.91182 \times 10^{-5}$& $1 + 2.91182 \times 10^{-5}$ & $-8.68202 \times 10^{-2}$ \\ 
		\hline
		$10^{-3}$  & $-2.91118 \times 10^{-7}$ & $1 + 2.91118 \times 10^{-7}$ & $-8.66989 \times 10^{-2}$\\ 
		\hline
		$10^{-4}$  & $-2.91117 \times 10^{-9}$ & $1 + 2.91117 \times 10^{-9}$ & $-8.66977 \times 10^{-2}$\\ [1ex] 
		\hline
	\end{tabular}
\end{center}

\begin{center}
	\begin{tabular}{||c | c | c | c||} 
		\hline
		$B^x/2$ & $\kappa_1$ & $\kappa_2$ &  $q_1$ \\ [0.5ex] 
		\hline\hline
		$10^{-1}$ & $-1.07577\times 10^{-3}$ & 
		$1 + 1.07577 \times 10^{-3}$ & $3.78262$\\ 
		\hline
		$10^{-2}$  & $-1.03663 \times 10^{-5}$& $1 + 1.03663 \times 10^{-5}$ & $3.77776$ \\ 
		\hline
		$10^{-3}$  & $-1.03624 \times 10^{-7}$ & $1 + 1.03624 \times 10^{-7}$ & $3.77771$\\ 
		\hline
		$10^{-4}$  & $-1.03623 \times 10^{9}$ & $1 + 1.03623 \times 10^{-9}$ & $3.77771$\\ [1ex] 
		\hline
	\end{tabular}
\end{center}

\bibliographystyle{plainurl} 
\bibliography{SecondPaperBibliography}

\end{document}